# The Effective Interfacial Tensions between Pure Liquids and Rough Solids: A Coarse-Grained Simulation Study

J. D. Hernández Velázquez[†,a], G. Sánchez–Balderas[b], A. Gama Goicochea[a], E. Pérez[‡,b]

[a] División de Ingeniería Química y Bioquímica, Tecnológico de Estudios Superiores de Ecatepec, 55210, Ecatepec de Morelos, Estado de México, México.

[b] Instituto de Física, Universidad Autónoma de San Luis Potosí, 78000, San Luis Potosí, SLP, Mexico.

## ABSTRACT

The *effective* solid-liquid interfacial tension (SL-IFT) between pure liquids and rough solid surfaces is studied through coarse-grained simulations. Using the dissipative particle dynamics method, we design solid-liquid interfaces, confining a pure liquid between two explicit solid surfaces with different roughness degrees. The roughness of the solid phase was characterized by Wenzel's roughness factor and the effective SL-IFT ($\gamma'_{sl}$) is reported as a function of it also. Two solid-liquid systems differentiated from each other by their solid-liquid repulsion strength are studied to measure the effects caused by the surface roughness on the calculation of $\gamma'_{sl}$. We found that the roughness produces changes in the structure of the liquid, which is observed in the first layer of liquid near the solid. These changes are responsible for the effective SL-IFT increase as surface roughness increases. Although there is a predominance of surface roughness in the calculation of $\gamma'_{sl}$, it is found that the effective SL-IFT is directly proportional to the magnitude of the solid-liquid repulsion strength. The insights provided by these simulations suggest that the increase of Wenzel's roughness factor is a direct consequence of the increase in surface area due to the vertical deviations measured in the topography. This, in turn, produces an increase in the number of effective solid-liquid interactions between particles, eventually yielding significant changes in the local values of the normal and tangential components of the pressure tensor.

# 1 Introduction

Surface wetting phenomena in a solid-liquid system involve also the surrounding media.[1] Understanding the intermolecular interactions between solids and liquids is crucial for technological advances in many applications, such as self-cleaning materials, liquid separation, corrosion protection,[2,3] and the science itself.[4,5] The wetting behavior of materials depends on their physical and chemical properties. It is known that a rough surface is more repellent than a flat surface made of the same material as the rough one.[5,6] A pioneering experiment that determined contact angles on rough surfaces was carried out by Johnson and Dettre in 1964.[6] They evidenced the influence of roughness on contact angles and its hysteresis, although their roughness measurements were qualitative. It is known that on a flat surface, the surface tensions define the wettability of the liquid on such a surface. However, roughness and structure play important roles, which means that the physical properties of the surface must be studied in detail. The solid-liquid interactions at the interface depend on the individual solid and liquid interactions. Dispersive (London) forces and non-dispersive (polar) forces are at play between the two media when they are in contact. The wettability of the surface depends strongly on van der Waals interactions, which are crucial in physical adsorption properties.[2]

The interfacial phenomena between a liquid phase and a solid phase have been studied for a long time. Young and Dupré[7] introduced the well-known relation between the contact angle and the interfacial tensions in a flat and homogeneous solid-liquid system, given by:

$$\gamma_{sl} = \gamma_{sv} - \gamma_{lv} \cos \theta_Y, \qquad (1)$$

where $\gamma_{sl}$, $\gamma_{sv}$, and $\gamma_{lv}$ are the surface tensions for solid-liquid, solid-vapor, and liquid-vapor involved in the wetting phenomenon, respectively, and $\theta_Y$ is Young's contact angle. Wenzel extended Young's equation for rough surfaces by introducing a roughness factor ($r_a$), defined as the ratio between the actual area of the surface and its in-plane projected area. He argued that for actual solid surfaces, the product $\gamma_{lv} \cos \theta_Y$ gauges not the difference $\gamma_{sv} - \gamma_{sl}$, but the product $r_a(\gamma_{sv} - \gamma_{sl})$, calling such modified value the "effective adhesion tension".[8,9]

Within the scope of numerical simulations, solid-liquid systems have been widely studied to gain further understanding of the mechanisms of wetting on surfaces. For instance, the kinetics of liquid spreading and wetting of fibers has been studied through the calculation of dynamic contact angles using molecular dynamics.[10] Despite the progress achieved with numerical methods, there are still several challenges remaining in this field. One of them is to determine the exact contribution of the deviations normal to the surface of the solid phase involved in the wetting phenomena. Specifically, an important issue is how the solid-liquid interfacial tension is affected by the roughness produced by such normal fluctuations measured from the topography of the solid phase. Many works have studied the effects of the structure of the solid phase on the calculation of the solid-liquid interfacial tension.[11–13] However, it is still challenging to understand how the changes in surface roughness, beyond geometrical patterns, modify the solid-liquid interfacial tension. For this purpose, we have carried out a set of simulations of solid-liquid interfaces to obtain the *effective* solid-liquid interfacial tension (SL-IFT, $\gamma'_{sl}$), for two types of pure liquids, which are different only by their repulsive interactions with the solid phase. The term "effective" means that in the calculation of the SL-IFT, the influence of the surface roughness is taken into account. This consideration is inspired by the contributions of Wenzel to the models of the wetting phenomena, arguing that the solid-liquid and solid-vapor interfacial tensions are magnified with increasing $r_a$.[9,14]

A frequent way to estimate the roughness of a surface is through its topographical data. Most of these data are obtained by microscopy techniques, such as atomic force microscopy, scanning electron microscopy and laser scanning confocal microscopy,[15–17]. Therefore, the roughness ($r_a$) of the solid surfaces designed in this work was calculated through their topographical data, by means of the matrix of heights. In general, we find that $\gamma'_{sl}$ increases as $r_a$ does. We note that the number of interactions between solid and liquid particles located in the interfacial region is responsible for the increasing values of $\gamma'_{sl}$. Additionally, the reason for the variations in the local values of the normal and tangential components of the pressure tensor is found to be due to the larger cavities formed as the surface roughness grows. The trends reported in this work are in agreement with experimental and numerical results in the literature, where the SL-IFT grows as the roughness of the surface.[16–18]

**2 Models and methods**

**2.1 Dissipative particle dynamics method**

In this work, we performed a set of simulations of two different pure fluids (liquid phase) confined by two fixed surfaces (solid phase) on the *z*-direction of the simulation box to model the solid-liquid interface, as seen in Fig. 1. These numerical models were solved using the dissipative particle dynamics (DPD) method,[19,20] a well-known and widely used technique for studying both the liquid-liquid interfacial tensions[21,22] and solid-liquid interfacial tension.[11,13]

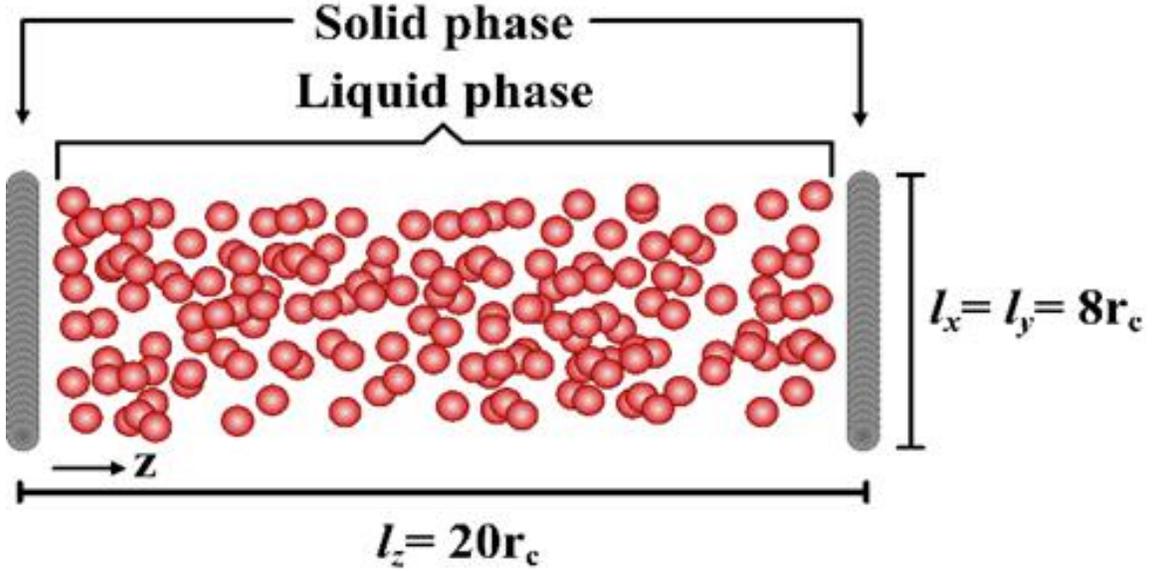

**Fig. 1** Sketch of the systems designed in this work. Both the solid and liquid phases are represented explicitly by red circles. The confinement is along the *z*-direction, which is perpendicular to the interface.

The DPD method involves three fundamental forces: a conservative one ($\mathbf{F}^C$), a random ($\mathbf{F}^R$) one, and a dissipative ($\mathbf{F}^D$) one. All the particles' ($i = 1, 2, \ldots, N$) positions $r_i$ and velocities $v_i$ in DPD, as in a conventional molecular dynamic, are governed by Newton's laws of motion. The total force acting on the *i*-th particle is the sum of the three fundamental forces:

$$\mathbf{f}_i = \sum_{i \neq j}^{N} \left[ \mathbf{F}_{ij}^C + \mathbf{F}_{ij}^R + \mathbf{F}_{ij}^D \right]. \tag{2}$$

These pair-wise forces depend on a finite cut-off radius $r_c^*$, which is the usual effective interaction range of all forces, and it is the intrinsic unit length of the DPD method ($r_c^* = 1$); beyond the cutoff radius, all the forces are equal to zero. The nature of $\mathbf{F}_{ij}^C$ is defined as a soft and linearly decaying repulsion:

$$\mathbf{F}_{ij}^C = a_{ij}\left(1 - r_{ij}^*/r_c^*\right)\hat{\mathbf{r}}_{ij}, \tag{3}$$

where $a_{ij}$ is the maximum value of the force between particles. The random ($\mathbf{F}_{ij}^R$) and dissipative force ($\mathbf{F}_{ij}^D$) have the same functional form than $\mathbf{F}_{ij}^C$, expressed as follows:

$$\mathbf{F}_{ij}^R = \sigma\left(1 - r_{ij}^*/r_c^*\right)^2 \zeta_{ij} \hat{\mathbf{r}}_{ij}, \tag{4}$$

$$\mathbf{F}_{ij}^D = -\gamma\left(1 - r_{ij}^*/r_c^*\right)\left(\mathbf{r}_{ij} \cdot \mathbf{v}_{ij}\right) \hat{\mathbf{r}}_{ij}, \tag{5}$$

where $\zeta_{ij} = \zeta_{ji}$ are random numbers with Gaussian properties. Here $\mathbf{r}_{ij}$ and $\mathbf{v}_{ij}$ are the relative position and velocity vectors, respectively, and $\hat{\mathbf{r}}_{ij}$ is the unit position vector. Random and dissipative forces are related through their strength values ($\sigma$ and $\gamma$), by the fluctuation-dissipation theorem,[19] given by the equation $k_B T^* = \sigma^2/2\gamma$, where $k_B$ is Boltzmann's constant, and $k_B T^*$ is the intrinsic unit energy of the DPD model.[19] The values of the random and dissipative force intensities were set in $\sigma = 3$ and $\gamma = 4.5$ to keep the value $k_B T^* = 1$.

All the simulations were carried out using a hybrid algorithm based on the implementation of the Monte Carlo (MC) method in the Grand Canonical (GC) ensemble, combined with DPD (ref AGG). The algorithm consists of 10 DPD steps per MC step, which allows us to create new velocities and positions for all the particles using the total force of the particles (eq. 2). Once the 10-step DPD dynamics is finished, the total energy of the system at the end is calculated and compared with the total energy before the 10-step DPD dynamics, and the configurations are accepted or rejected according to the Metropolis criterion.[23] Once the new configuration has been accepted or rejected after the MC cycle, the algorithm proceeds to the exchange of liquid particles with their bulk phase to keep the chemical potential ($\mu$) constant, according to the selection rules of the GC ensemble. The hybrid GCMC-DPD algorithm has been used successfully in the study of thermodynamic properties of confined fluids, colloidal stability, polyelectrolyte adsorption, etc.[24–26] Each simulation was run in 20 blocks of $10^4$ MC steps, with the time step of $\Delta t = 0.03\tau$, with $\tau \approx 3\ ps$. The first 10 blocks were used for the equilibrium phase, and the last 10 blocks were used for the production phase. All the simulations were performed using reduced units, and the intensities of the conservative force between the same particle species were chosen as $a_{ii} = 78.3$. This choice corresponds to a coarse-graining degree of three, which is the volume of three water molecules grouped into a DPD particle.[27] Two types of pure liquids interacting with four different surfaces were studied in this work, yielding a total of eight different solid-liquid systems. The values of the cross-interaction parameters ($a_{ij}$) of the conservative DPD force between solid and liquid particles were chosen as $a_{ij} = 83.5$ and $a_{ij} = 140$ units. Larger values of the $a_{ij}$ parameter produce larger repulsion, between the solid and liquid species.

The solid surfaces were built by "freezing" the degrees of freedom of the DPD particles that make them up. Care was taken to remove those degrees of freedom when the kinetic energy of the systems is calculated. We opted for the generation of the surfaces as random, homogeneous solids to avoid implementing geometric structures and patterns. Several studies have shown that both geometric structure and patterns on the surfaces influence their wetting properties.[11,12,28–30] Fig. 2 shows the solid surfaces modeled in this study.

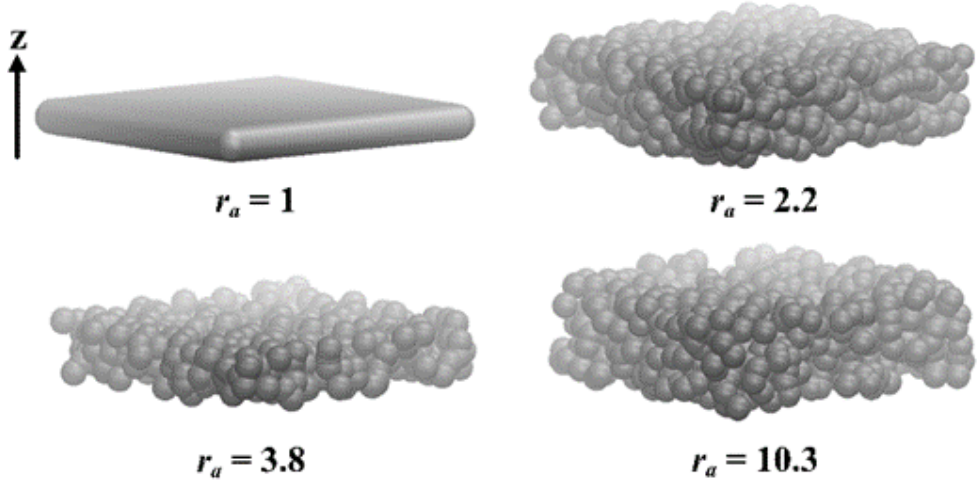

**Fig. 2** Snapshots of the solid surfaces modeled in this work with different roughness degrees, measured with the roughness factor, $r_a$. See eq. (6).

## 2.2 Wenzel's roughness factor

The quantification of the roughness degree of our surfaces was made by the roughness factor ($r_a$) proposed by Wenzel, which is defined as the ratio of the so-called real area ($A_R$) and the geometric area ($A_G$):[9]

$$r_a = A_R/A_G, \tag{6}$$

where the $A_G$ is the in-plane projected area, perpendicular to the confinement direction ($l_z$ in this case). $A_R$ is the total area of the surface, which is approximated with a triangular tessellation method. Thus, $r_a = 1$ is the value for an ideal flat surface. The key aspect of $r_a$ is to calculate $A_R$. That is because the more accurately $A_R$ is obtained, the more accurately the $r_a$ value is. Moreover, it is known that $A_R$ depends on the scan size of the surface, the resolution (i.e., pixel number) of the image, and the tessellation method.[31,32] In this work, the $r_a$ values were calculated for images with lateral sizes of $l_x = l_y = 8\, r_c$ and a resolution of $1024 \times 1024$ pixels. The geometric area of the images is $A_G = 64\, r_c^2$, while $A_R$ was obtained by using a triangular tessellation method. $A_R$ is the average value of a clockwise and a counterclockwise triangulation, considering each pixel $(p_{i,j})$ of the image as a vertex of a triangle. Thus, the approximated real area is given by the following equation:

$$A_R = \sum_{i=1}^{n_x=N} \sum_{j=1}^{n_y=N} \frac{1}{2} \left\{ \left[ \frac{|(p_{i,j}, p_{i+1,j})||(p_{i,j}, p_{i,j+1})|}{2} + \frac{|(p_{i,j+1}, p_{i+1,j+1})||(p_{i+1,j}, p_{i+1,j+1})|}{2} \right] \right.$$

$$+ \left[ \frac{|(p_{i,j}, p_{i,j+1})||(p_{i,j+1}, p_{i+1,j+1})|}{2} \right.$$

$$\left. \left. + \frac{|(p_{i,j}, p_{i+1,j})||(p_{i+1,j}, p_{i+1,j+1})|}{2} \right] \right\}, \qquad (7)$$

where the distances are represented by the vertical bars, | |. The first term within square brackets on the right-hand side of equation (7) is the approximated $A_R$ of the clockwise triangles. The second one is for the counterclockwise triangles (where $i = 1, 2, \ldots, n_x$ and $j = 1, 2, \ldots, n_y$ are the number of pixels in the x- and y-directions, respectively); and $n_x = n_y = 1024$, is the number of pixels in the *x*- and *y*-directions, respectively. Fig. 3 shows a schematic representation of the clockwise method used to calculate the real area of the images.

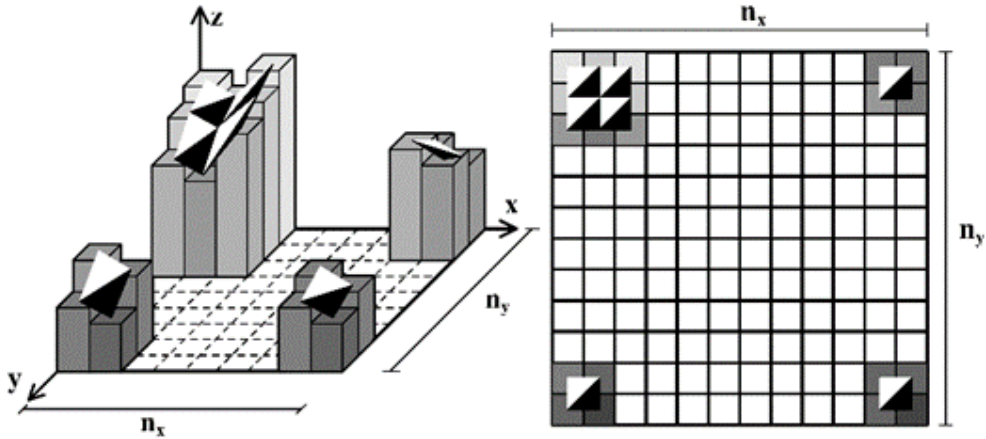

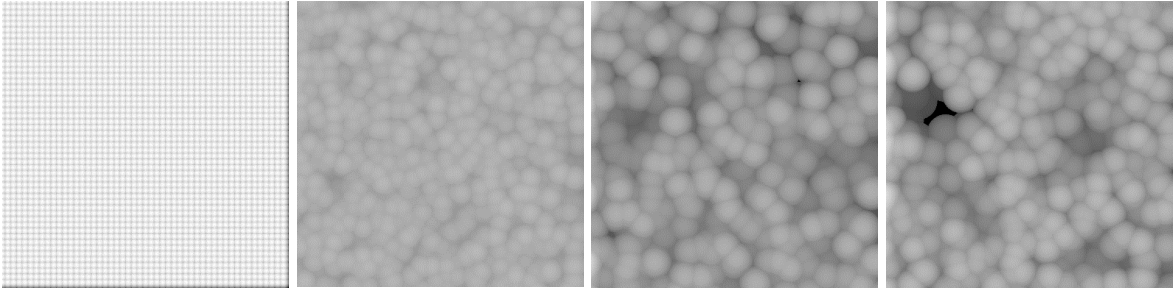

**Fig. 3** (*Top*) Illustration that shows the clockwise tessellation method applied to approximate the real area of the grayscale images of resolution $n_x \times n_y$. (*Bottom*) Grayscale images of the topographies from which the surface roughness factor ($r_a$) was calculated, from smoother to rougher surfaces ($r_a = 1$ to $r_a = 10.3$, from left to right). In this work, $r_a$ was calculated using the resolution of the images as $N = n_x = n_y = 1024$ pixels and the lateral size is $L = l_x = l_y = 8\, r_c$.

## 3 Results and Discussion

In the density profiles presented in Fig. 4 one can observe the average structure that the system has. It is important to calculate those profiles because it has been shown that the structure of confined simple fluids is related to most of their mechanical properties thereof.[13] That being the case, let us start with the analysis of the reduced density ($\rho^*$) profiles shown in Fig. 4. The first relevant aspect of all of them is the layering of the liquid phase near the solid phase, where the gray vertical lines in Fig. 4 represent the inner edge of the solid surface. This layering is expected when modeling confined simple or complex fluids by explicit or effective (implicit) parallel walls.[13,33,34] We note that the number of layers, indicated by the maxima near the solid phase, is affected neither by increasing roughness nor by varying the intensity of the repulsive force $a_{ij}$. This is a direct consequence of the random nature of the surfaces, lacking geometric patterns. Those have been shown to disappear and/or modulate the layering of confined simple fluids.[13] In addition, the periodicity shown near the solid phases of all the density profiles is similar and it is approximately equal to the diameter of the DPD particle.[34] However, the liquid with high repulsion force ($a_{ij} = 140$, red curves in Fig. 4) always forms the first layer with higher number of particles. This is expected from the fact that, if the liquid and solid phases are more dissimilar (i.e., larger values of $a_{ij}$), the liquid will tend to avoid closer contact with the solid, forming a denser layer of fluid particles. This is in contrast with fluids that possess lower values of $a_{ij}$. These results are expected only when non-dispersion forces are absent, i. e., when the nature of the intermolecular forces is purely dispersion forces, as in these systems.

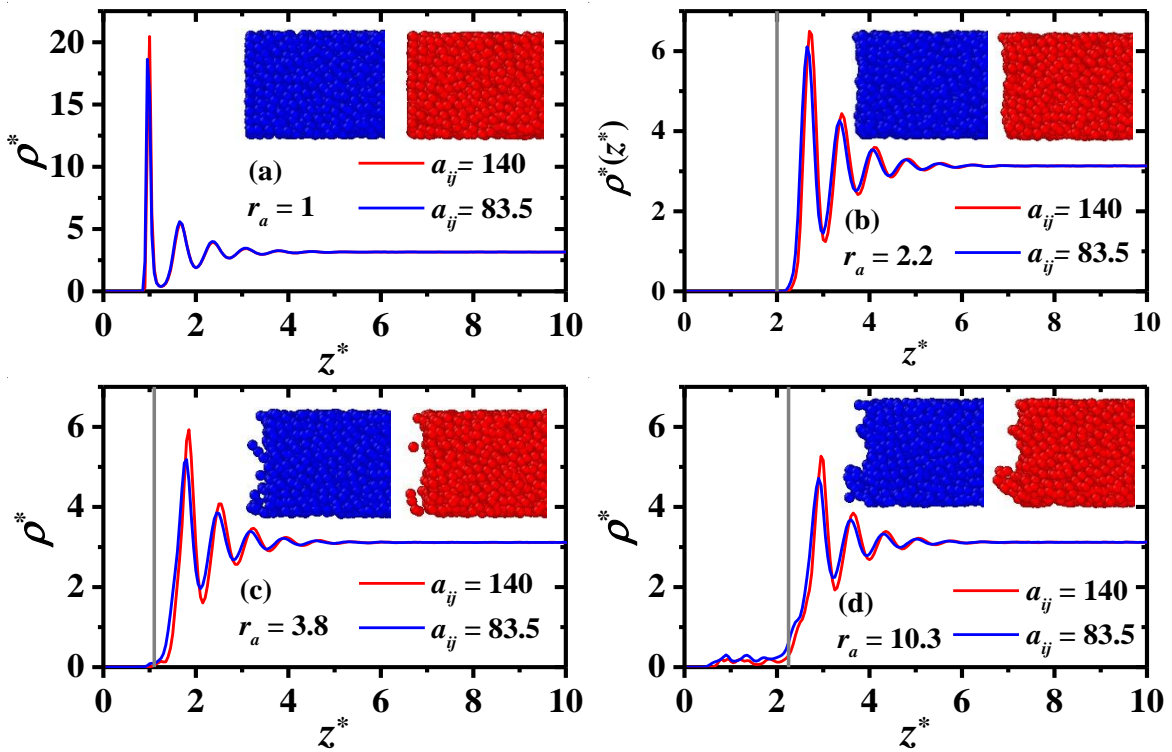

**Fig. 4** Density profiles of both fluids with high and low repulsion force parameters (red and blue curves, respectively) interacting with each of the four different rough surfaces. The profiles are obtained along the confinement direction ($z^*$). A snapshot of both liquids is included within each panel to show the section of fluid that penetrates the solid surface (solid phase omitted for clarity). The gray vertical lines in all the panels mark the inner edge of the surface. Asterisked quantities are expressed in reduced DPD units.

The surface roughness barely affects the layer ordering of the liquid structure. This is seen in the density of the first layer near the solid phase, which decreases by ∼ 23% and ∼ 19% for liquids of $a_{ij} = 83.5$ and $a_{ij} = 140$, respectively, as the surface roughness factor increases from $r_a = 2.2$ to $r_a = 10.3$; see Fig. 4. However, a salient aspect is that, in the transition from "smooth" ($r_a = 1$) to "rough" (in this case $r_a = 2.2$) surfaces, the difference between the densities of the first layer is reduced by ~ 67% and ~ 68% for liquids of $a_{ij} = 83.5$ and $a_{ij} = 140$, respectively. Furthermore, we note that as the roughness grows, the first peak in the density profiles of the liquids with $a_{ij} = 140$ exhibits a slight shift toward the bulk direction. That is the zone within the simulation box where the solid phase does not influence the fluid's particles and the density remains constant ($\rho^* = 3$). This also is a consequence of the dissimilarity between the liquid and solid phase because the increase in the surface roughness must produce an increase in the interfacial region. Thus, one would expect the liquid layers to be displaced proportionally to the characteristic intensity of the repulsion forces between the two phases.

The predominant role of surface roughness, whose main characteristics are its randomness and homogeneity, is that it allows the penetration of fluid's particles into the solid phase as the surface becomes rougher, since it has larger and larger cavities as the roughness degree rises, as can be seen in Fig. 4. In addition, it should be noted that the surface capacity to absorb the liquid depends on the dissimilarity of the interactions between the fluid and the solid particles. For example, when this dissimilarity is low ($a_{ij} = 83.5$, blue lines in Fig.4) the amount of liquid that penetrates the solid phase is higher than that when the soli-liquid dissimilarity is large, independently of the surface roughness. Such mechanism lays the foundation for the results presented below, which are directly linked to changes in the effective SL-IFT.

To calculate the effective SL-IFT, we use the Irving-Kirkwood (IK) definition, which states that the contributions to the stress across a surface element $dA$ are due to the pairs of particles whose position vector $r_{ij} = r_j - r_i$ crosses said $dA$.[35] The IK approach is widely used in the DPD method to compute the interfacial tensions of different types of solid-liquid and liquid-liquid interfaces, including interfaces formed with organic solvents or complex fluids.[13,21,22,36–38] In the IK approach, the interfacial tension is calculated by integrating the difference between the normal and tangential components of the stress as follows:[35]

$$\gamma^*(z^*) = \int_0^{l_z} [P_N^*(z^*) - P_T^*(z^*)]dz^*, \tag{8}$$

where $l_z$ is the box–size in the $z$-direction. The normal and tangential components of the pressure tensor are defined as $P_N^*(z^*) = P_{zz}(z^*)$ and $P_T^*(z^*) = 1/2\left[P_{xx}(z^*) + P_{yy}(z^*)\right]$, respectively. The components of the pressure tensor are obtained from the virial route (ref). The asymmetry in the pressure tensor **P** is caused by the confinement along the $z$-direction, therefore the normal component of **P** must be written differently from the tangential component. From this definition, it is

(AQUÍ TE FALTÓ PONER LA ECUACIÓN)

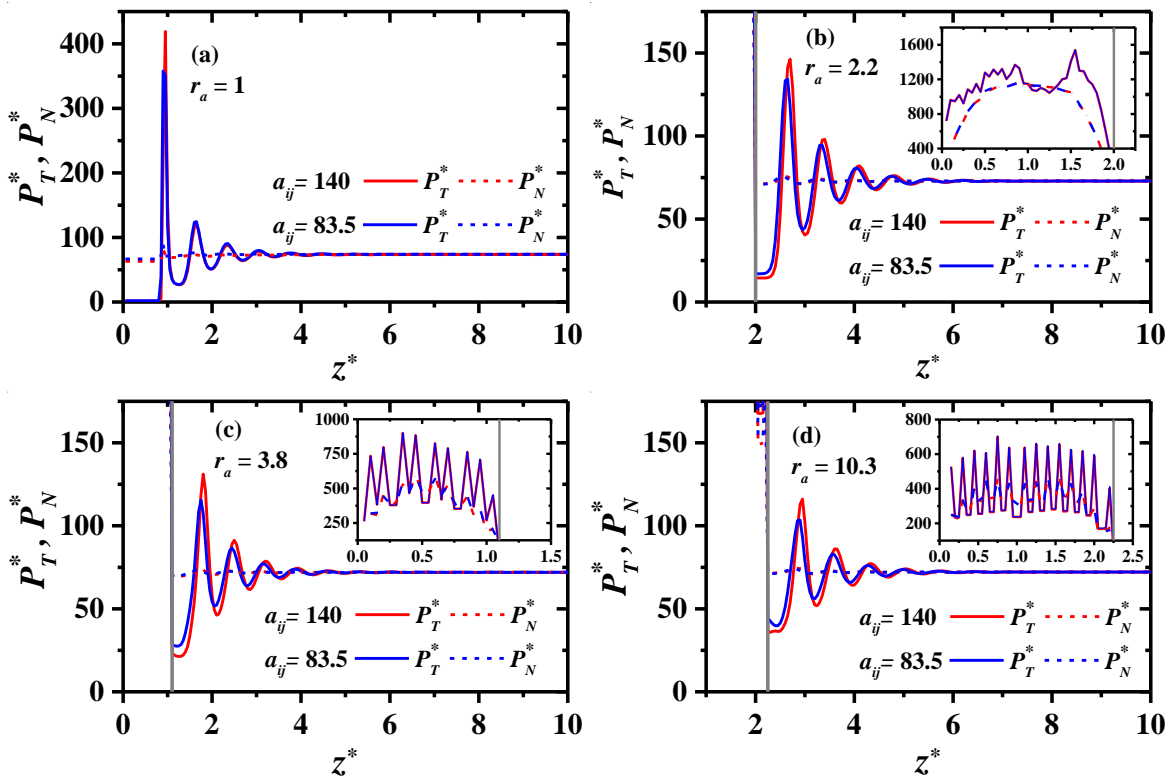

**Fig. 1** Profiles of the tangential ($P_T^*$, continuous lines) and normal ($P_N^*$, dashed lines) components of the pressure tensor along the confinement direction for the systems with high (red lines) and low (blue lines) solid-liquid dissimilarity. The insets show the local values of $P_T^*$ and $P_N^*$ within the solid phase. The vertical grey lines in both main panels and insets of figures (b), (c), and (d) indicate the inner edger of the solid surface such as in Fig. 4. Asterisked symbols represent quantities in reduced DPD units.

In Fig. 5 we show the profiles of the tangential ($P_T^*(z^*)$, continuous lines), and normal ($P_N^*(z^*)$, dashed lines) components of the pressure tensor along the confinement direction, $z$. We find that the $P_T^*$ profiles exhibit the same oscillatory behavior as that seen in the density profiles (see Fig. 4). The $P_T^*$ profiles of systems with $a_{ij} = 140$ clearly show higher amplitude of the maxima, as occurs in the density profiles of the liquids with the same interactions. This is to be expected, since larger repulsion ($a_{ij}$) between the solid and liquid phases leads to higher IFT. We note that the normal pressure, $P_N^*$, remains constant within the region confined by the explicit solid walls, marked by the grey vertical lines in Fig. 5. This qualitative trend indicates that the confined fluid is in mechanical equilibrium, as argued by Terrón-Mejía et. Al.[13] They worked with liquids confined by sinusoidal implicit walls, represented by an effective force instead of building them with explicit DPD particles, as done in this work. The normal and tangential components of the pressure in the region where the liquid is adsorbed on the solid phase (see insets in Fig. 5) show that the local values of the pressure are considerably higher than those of the interfacial, because of the contribution to the pressure tensor due to the solid-solid interactions. (Esta oración no se entiende y está muy larga)

Figure 6 shows the solid-liquid interfacial tension profile (SL-IFT) profiles of the systems, as a cumulative sum of the absolute differences between $P_N^*$ and $P_T^*$, ($\gamma_{sl}^*(z^*) = |P_N^* - P_T^*|$), along the confinement direction. Presenting the IFT results in this way makes it easier to identify where the interface and bulk regions are located. The former can be found wherever the IFT profile shows a positive slope, while the latter is found wherever a plateau region exists. Here is important to remark that the SL-IFT is calculated from the $z^*$ value where the interfacial region starts. Such region is defined as the solid-liquid coexistence region, which is obtained from the density profiles (Fig. 4). For rough surfaces, it starts from the $z^*$-point where the liquid density profile starts.

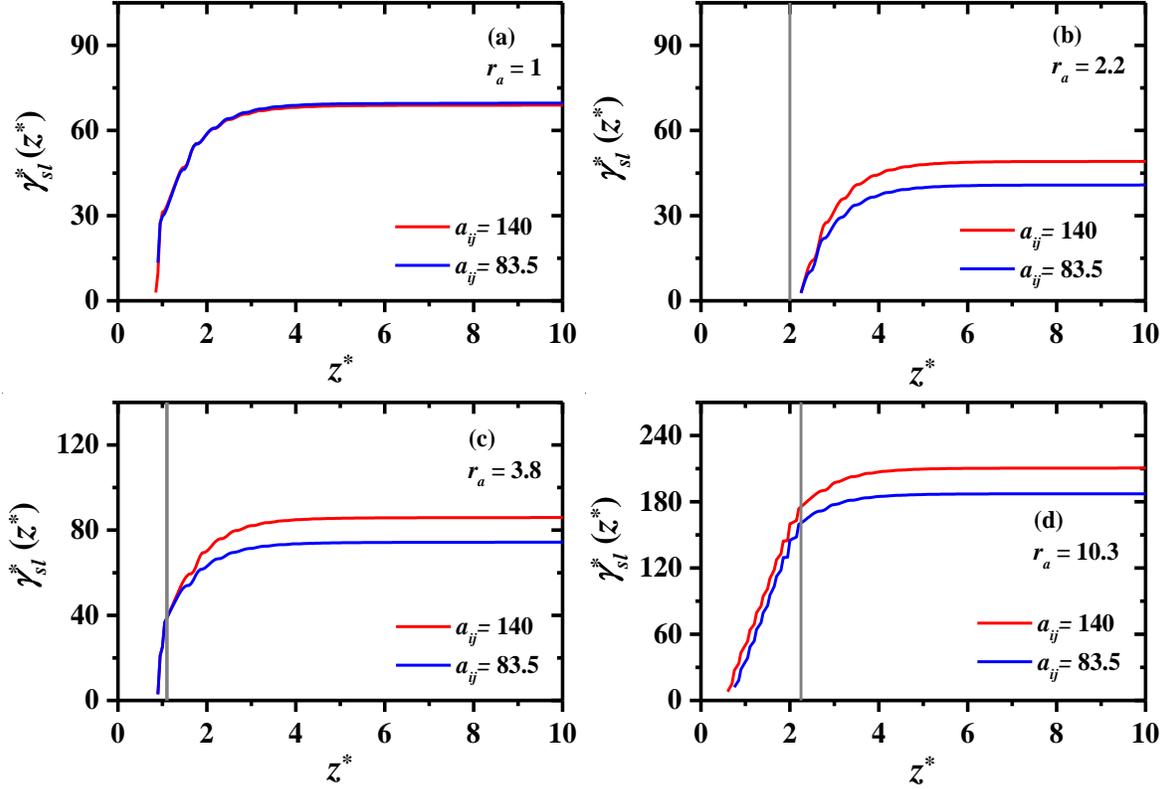

**Fig. 2** Solid-liquid interfacial tension profiles of systems with low (blue lines) and high (red lines) solid-liquid repulsion force, for surfaces with increasing roughness, $r_a$. Vertical gray lines indicate the inner edges of the solid phases, as in Figs. 4 and 5. Asterisks refer to reduced DPD units.

The results displayed in Figs. 4-6 help identify the wetting mechanism on rough surfaces because they show that the cumulative value of the difference between $P_N^*(z^*)$ and $P_T^*(z^*)$, i.e., the interfacial tension, grows as the roughness grows. This occurs despite the fact that the individual values of $P_N^*$ and $P_T^*$ are reduced with increased surface roughness; see Fig. 5. Therefore, if the interfacial region becomes thicker with increased roughness, then the solid-liquid interfacial tension $\gamma_{sl}^*$ is expected to grow as well; see eq. (8). That is because the value of the interfacial tension profiles at the end of the box ($z^* = 10$) is the result of the changes in the local values of the components of the pressure tensor along the interfacial region.

Using the IK definition (eq. 8) in the SL-IFT profiles of Fig. 6, we calculate the effective SL-IFT. Hereafter, we use the notation $\gamma'_{sl}$ to refer to the effective SL-IFT, which is obtained from the value of the plateaus in Fig. 6., It has been shown that in liquids confined by symmetric walls $\gamma'_{sl}$ is independent of the direction in which the integration in eq. (8) is carried out.[13] Wenzel suggests[9] that the effective SL-ITF can be taken as $\gamma'_{sl} \propto r_a \gamma_{sl}$ (cual es la diferencia entre $\gamma'_{sl}$ y $\gamma_{sl}$) in a first approximation, and our results agree with this suggestion. (Pero por qué? Cómo puede ver el lector que esta sugerencia se cumple en este trabajo?)

Lastly, Fig. 7 shows $\gamma'_{sl}$ as a function of the surface roughness factor $r_a$, which ranges from $r_a = 1$ up to $r_a = 10.3$. We observe that $\gamma'_{sl}$ grows for $r_a \geq 2.2$, independently of the dissimilarity between the solid and liquid phases. Although it is found that $\gamma'_{sl}$ does depend on the solid-liquid chemical dissimilarity (the repulsion parameter $a_{ij}$), such that the interfacial tension increase with the solid-liquid repulsion, the results reveal that surface roughness is the leading variable affecting $\gamma'_{sl}$. The trends we obtain for $\gamma'_{sl}$ as a function of $r_a$, agree with the results reported by Mittal and Hummer,[18] who study the effect of the surface roughness on the interfacial tension for water confined by a planar wall and a sinusoidal one. They show that by increasing the amplitude of the sinusoidal surface (i. e., the surface roughness), the character of the surface changes from hydrophilic to hydrophobic. In comparison with our results, we note that in the transition from smooth ($r_a = 1$) to rough ($r_a = 2.2$) surfaces, the wettability of the surface is enhanced, yielding a minimum in $\gamma'_{sl}$ when $r_a = 2.2$ for both solid-liquid systems. By continuing to increase the roughness, surface wettability worsens, thereby producing higher $\gamma'_{sl}$.

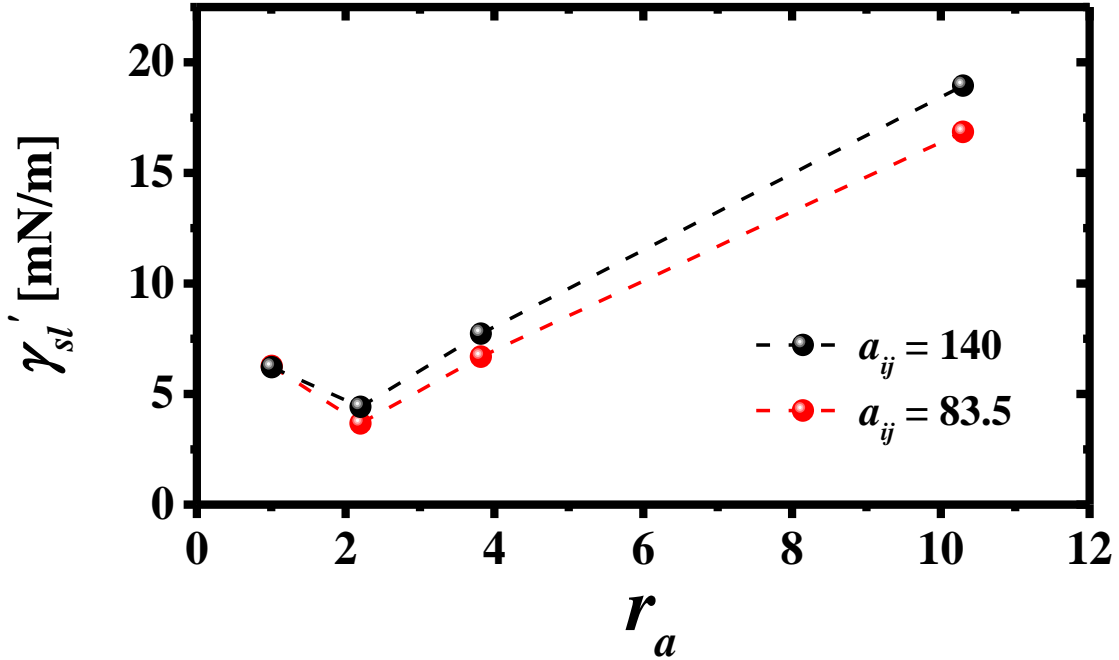

**Fig. 7** Effective solid-liquid interfacial tension ($\gamma'_{sl}$) as a function of the Wenzel roughness factor ($r_a$). Dashed lines are only guidelines.

**4 Conclusions**

In this work, the effective SL-IFT ($\gamma'_{sl}$) between pure liquids and rough surfaces is investigated using coarse-grained numerical simulations. A hybrid Monte Carlo-DPD method is used to model solid-liquid interfaces, where two different liquids interact with four different rough solids. The surface roughness was quantified with Wenzel's roughness factor ($r_a$), and the $\gamma'_{sl}$ were studied as a function of $r_a$. We find that the formation of cavities in the solid phase increases for rougher surfaces, and it is responsible for the increase of $\gamma'_{sl}$. The insights provided by our results indicate that the increase in the interfacial tension is due to fact that the local values of the pressure tensor in the interfacial region, are larger because of the growing number of solid-liquid interactions in the interfacial region The width of such interfacial region increases as the surface roughness does. These results show that better wettability occurs on smoother surfaces, in agreement with observations reported in the literature on hydrophobic surfaces.[16,17,39] In fact, in the transition from smooth ($r_a = 1$) to rough ($r_a = 2.2$) surfaces there appears an enhancement on the wettability of the surface, yielding a minimum in $\gamma'_{sl}$. This conclusion agrees with the numerical results provided by Mittal and Humer,[18] who noted that there is a hydrophilic-to-hydrophobic transition in the nature of the surface as roughness is increased. These results suggest that the surface wettability can be controlled by the modification of the surface roughness. Also, these results are expected to be useful for research related to the study of SL-IFT between rough surfaces and pure, non-polar liquids, and in interfacial phenomena where wetting properties must be modulated by the topography of the surfaces.

**Conflicts of interest**

"There are no conflicts to declare".

**Acknowledgments**

The authors thank J. Ireta and the *Laboratorio Nacional de Cómputo de Alto Desempeño*, at UAMI for granting us use of the computational resources. J. D. H. V. is particularly grateful to the late R. López-Rendón for his help over the years and to CONACYT for postdoctoral support. A.G.G. thanks CONACYT for support also, through grant 320197.